# THE FNAL INJECTOR UPGRADE*

C.Y. Tan[#], D.S. Bollinger, K.L. Duel, J.R. Lackey, W.A. Pellico, FNAL, Batavia, IL 60510, U.S.A.

*Abstract*

The present FNAL H- injector has been operational since the 1970s and consists of two magnetron H- sources and two 750 keV Cockcroft-Walton Accelerators. In the upgrade, both slit-type magnetron sources will be replaced with circular aperture sources, and the Cockcroft-Waltons with a 200 MHz RFQ (radio frequency quadrupole). Operational experience at BNL (Brookhaven National Laboratory) has shown that the upgraded source and RFQ will be more reliable, improve beam quality and require less manpower than the present system.

## INTRODUCTION

The present FNAL (Fermi National Accelerator Laboratory) injector has been operational since 1978 and has been a reliable source of H- beams for the Fermilab program. At present there are two Cockcroft-Walton injectors, each with a magnetron H- source with a slit aperture [1]. With these two sources in operation, the injector has a reliability of better than 97%. However, issues with maintenance, equipment obsolescence, increased beam quality demands and retirement of critical personnel, have made it more difficult for the continued reliable running of the H- injector. The recent past has also seen an increase in both downtime and source output issues. With these problems coming to the forefront, a new 750 keV injector is being built to replace the present system. The new system will be similar to the one at BNL (Brookhaven National Laboratory) that has a similar magnetron source with a round aperture and a 200MHz RFQ. This combination has been shown to operate extremely reliably [2].

## THE NEW INJECTOR

The new design can be divided into three parts: the source, the LEBT (low energy beam transport) and the MEBT (medium energy beam transport). The LEBT is the transport line before the RFQ and the MEBT is the transport line from the end of the RFQ to the beginning of the DTL (drift tube linac). A preliminary drawing of the new injector is shown in Figure 1.

### Magnetron Source

The design will contain two H- magnetron sources for increased reliability. Each H- magnetron source will be the round type and will be mounted on a slide. The beam out of the source is at 35 keV and should be >60 mA. The source design is nearly identical to the BNL round source [2].

The round aperture magnetron takes advantage of a spherical dimple in the cathode that focuses the H- produced at the anode opening. Also, the extraction scheme used by BNL has the extraction electrode connected to ground, and the source is pulsed to 35 kV. This means that the extraction voltage is the acceleration voltage. This is different than the existing system, where the extraction electrode is between the source and ground. The typical extraction voltage is 15 kV to 18 kV. An additional advantage of the round aperture is the reduced surface area and amount of cesium required to operate.

With the high extraction voltage, low arc current and spherical focusing the power efficiency of the BNL style source is about 67 mA/kW with a lifetime that is between 6 to 9 months [2].

Contrast this to the present source which needs an arc current of 45 A with an arc voltage of 140 V to provide H- beam current of 50 mA. This gives a power efficiency of about 8 mA/kW. This relatively low power efficiency contributes to source aging caused by erosion of the cathode material. The lifetime of the source is on average 3.5 months.

### LEBT

The LEBT optics is a standard one where two solenoids are separated by a short distance so that the beam at the source and at the entrance of the RFQ are at the focal points of each solenoid. The beam is space charge dominated and Xe gas will also be used for neutralizing and focusing of the H- beam because it has been shown at BNL that there is increased transmission efficiency when Xe gas is used [3].

An Einzel lens installed near the entrance of the RFQ will be used as the chopper because it is much easier to chop the beam at low energy and also there is insufficient space in the MEBT. It is necessary to place the Einzel lens very close to the RFQ to minimize the area where beam will be de-neutralized. A pure electrostatic kicker will de-neutralize the H- and any advantage of Xe gas focusing will be lost during the chopping process [4]. Experiments at FNAL using a very similar Einzel lens have shown that it works well as a chopper and the beam pulse will have a rise time of < 50 ns and is mainly dominated by the pulser rise time.

### RFQ

The FNAL RFQ has been ordered from A. Schempp (Institut für Angewandte Physik). A solid CAD model of the RFQ is shown in Figure 3 and some of its physical and operating specifications are shown in Table 1. The transmission efficiency of the RFQ is 99.7% from the PARMTEQM simulation. The result is shown in Figure 5.

___________________________________________
*Work supported by DOE contract no. DE-AC02-07CH11359
[#]cytan@fnal.gov

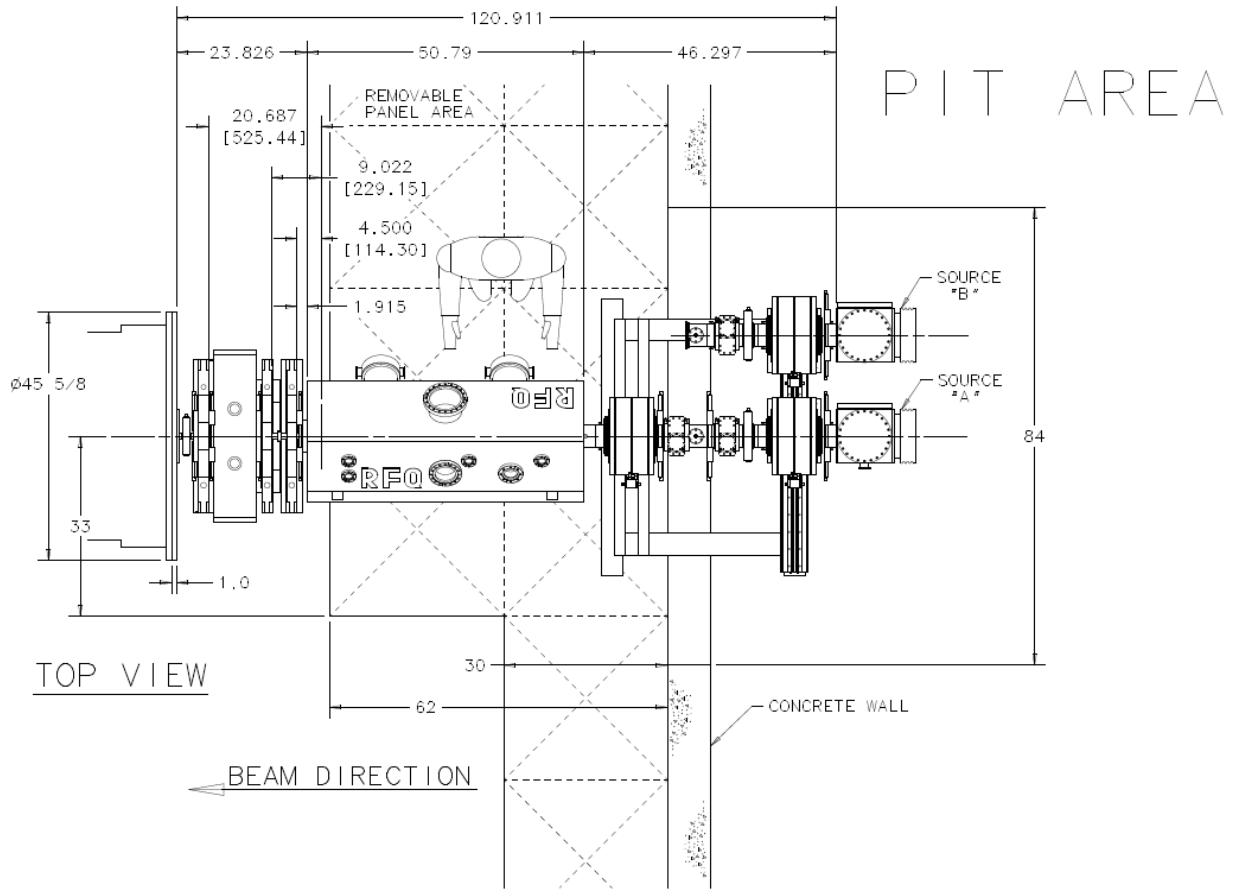

Figure 1: A preliminary drawing of the new injector connected to Tank 1. Shown here are the 2 H- sources for redundancy, the LEBT, RFQ and MEBT. Need to be updated with new MEBT.

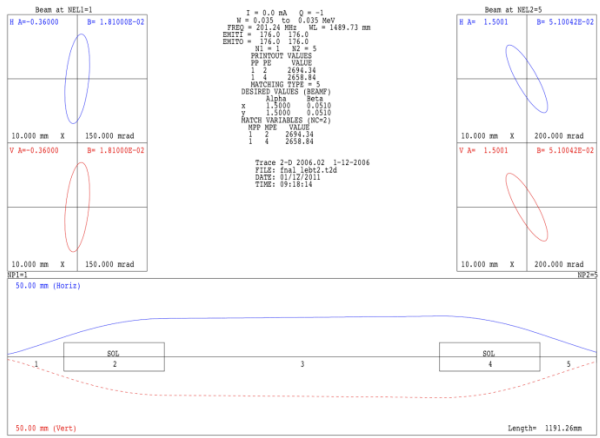

Figure 2: Trace2D model of the LEBT for zero current H- beam from the source to the entrance of the RFQ.

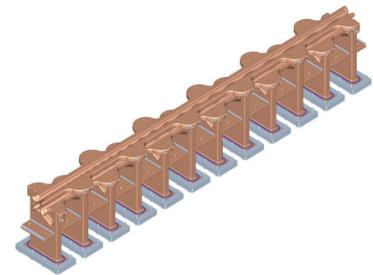

Figure 3: The solid engineering model of the RFQ.

## MEBT

The MEBT lattice is quadrupole doublet – buncher – quadrupole doublet. The beam is essentially round at the output of the RFQ. The bunched beam out of the RFQ is space charge dominated and puts significant constraints on the MEBT design to prevent both longitudinal and transverse emittance blow up. The MEBT requirements on length and lattice required extensive design work for the beam line elements. PARMILA simulations with the final design settings, shown in Figure 4, show that at

60mA, 95.1% of the H- beam is captured at the end of DTL 1.

The main challenge in this design is the quadrupoles that have to be thin and run with a very high gradient. FNAL's Technical Division has produced a design that has a physical length of 2.6" and has corrector dipoles embedded.

The MEBT also requires one buncher. The buncher design is from BNL which was recently (end of 2010) installed in the BNL injector. The FNAL buncher was delivered recently and is in the process of being vacuum certified and low power tested.

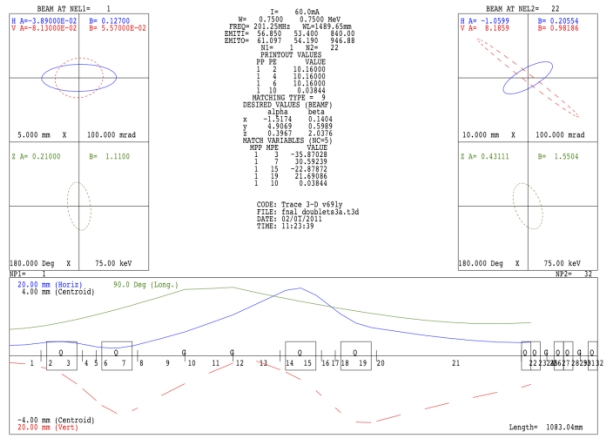

Figure 4: Trace3D model of the MEBT from the exit of the RFQ to the start of the DTL.

## CONCLUSION

The injector upgrade is proceeding as planned. Most of the parts are being manufactured and will arrive for testing before the end of 2011. It is slated for installation in 2012. The efforts of many at FNAL, BNL and elsewhere have helped to accomplish the fast pace of the project.

## ACKNOWLEDGEMENTS

We would like to thank the BNL pre-injector systems group for generously helping us in the design of the injector.

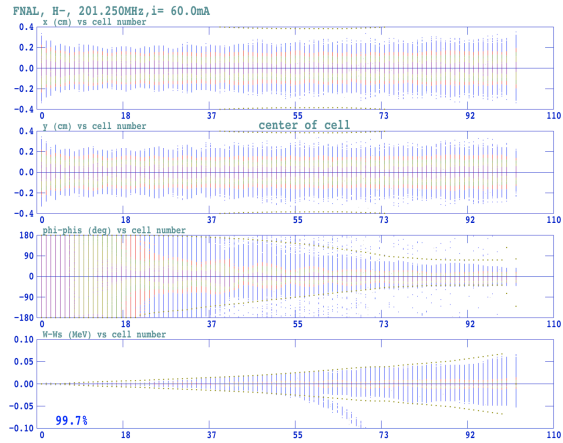

Figure 5: PARMTEQM simulation of the RFQ which shows the transmission efficiency is 99.7%.

Table 1: RFQ Specifications

| Parameter | Value | Units |
| --- | --- | --- |
| Input Energy | 35 | keV |
| Output Energy | 750 | keV |
| Frequency | 201.25 | MHz |
| Num. of cells | 102 | |
| Length | 120 | cm |
| Duty factor | 0.12 | % |
| Design current | 60 | mA |
| Peak power | < 100 | kW |